# Estimating achievement from fame[1]

M.V. Simkin and V.P. Roychowdhury

*NetSeer Inc. - 11943 Montana Ave. Suite 200, Los Angeles, CA 90049 and Department of Electrical Engineering, University of California, Los Angeles, CA 90095-1594*

We report a method for estimating people's achievement based on their fame. Earlier we discovered that fame of fighter pilot aces (measured as number of Google hits) grows exponentially with their achievement (number of victories). We hypothesize that the same functional relation between achievement and fame holds for other professions. This allows us to estimate achievement for professions where an unquestionable and universally accepted measure of achievement does not exist. We apply the method to Nobel Prize winners in Physics. For example, we obtain that Paul Dirac, who is hundred times less famous than Einstein contributed to physics only two times less. We compare our results with Landau's ranking.

Earlier we discovered [1] that fame of WWI fighter pilot aces (measured in Google hits) grows exponentially with their achievement (measured in victories). Since then Bagrow *et al* found [2], that for physicists the relation between achievement and fame is linear. The measure of achievement used in that study was the number of published papers. However, Bogdanoff affair [3] demonstrated that one could publish in respectable journals even papers consisting of an incoherent stream of buzzwords of modern physics. Thus, we cannot use the number of published papers to measure scientific achievement. Garfield suggested [4] that the number of citations to scientist's papers is the true measure of scientific achievement. In another study [5], [6] we had shown that since citations multiplicate by mere copying this measure is also questionable,. While the number of citations may be increasing with the size of scientific contribution made in the paper, it is not obvious what the exact relation between these variables is. In this paper, we hypothesize that the same exponential relation between fame and achievement, as we found for fighter pilots, holds for people of other professions. We then use their fame (measured in Google hits) to infer their achievement[2].

In Ref.[1] we found that fame, $F$, depends on achievement, $A$, according to the following equation:

$$F(A) = C \times \exp(\beta \times A) \qquad (1)$$

Here $\beta$ and $C$ are parameters determined by regression. To be precise, the real data of fame as a function of achievement present not a smooth curve, but a scatter plot (see Fig 1 of [1]). Nonetheless, given the value of achievement, we can greatly reduce the uncertainty in the value of fame. Similarly, given the value of fame, we can try to estimate achievement. This we can do by simple inversion of Eq. (1):

$$A(F) = \ln(F/C)/\beta \qquad (2)$$

We will first see how it works using the aces data where we do know both fame and achievement. We computed for every ace an estimate of achievement based on his fame using Eq.(2). We then divided it by his real achievement. Figure 1 shows the distribution of such ratios of estimated and real achievements for 392 German WWI aces studied in Ref [1]. With high accuracy, we can approximate it by a lognormal distribution with mean zero and variance of 0.49. Kolmogorov-Smirnov test is passed with the p-value of 0.40. Analysis of the data of Fig.1 shows that with 50% probability estimated achievement is between 0.7 and 1.44 of real achievement. With 95% probability, estimated achievement is between 0.43 and 2.4 of real achievement. And with the 85% probability the real achievement is between two times more and two times less than the estimate. The estimate is thus not very accurate; however, even such crude an estimate can provide some insight in the fields where we have no clue of how to measure achievement.

---
[1] A modified version of this article under the title "Von Richthofen, Einstein and the AGA" was published in March 2011 issue of Significance http://www.significancemagazine.org/details/magazine/1036513/Von-Richthofen-Einstein-and-the-AGA-.html

[2] We do not insist that web hit counts are preferable to citation counts. These two measures of fame are strongly correlated and are interchangeable. We used web hits because we used them for fighter pilots aces in our earlier study. The point of this paper is not that one should use web hits, but that one should take a logarithm of fame to estimate achievement.

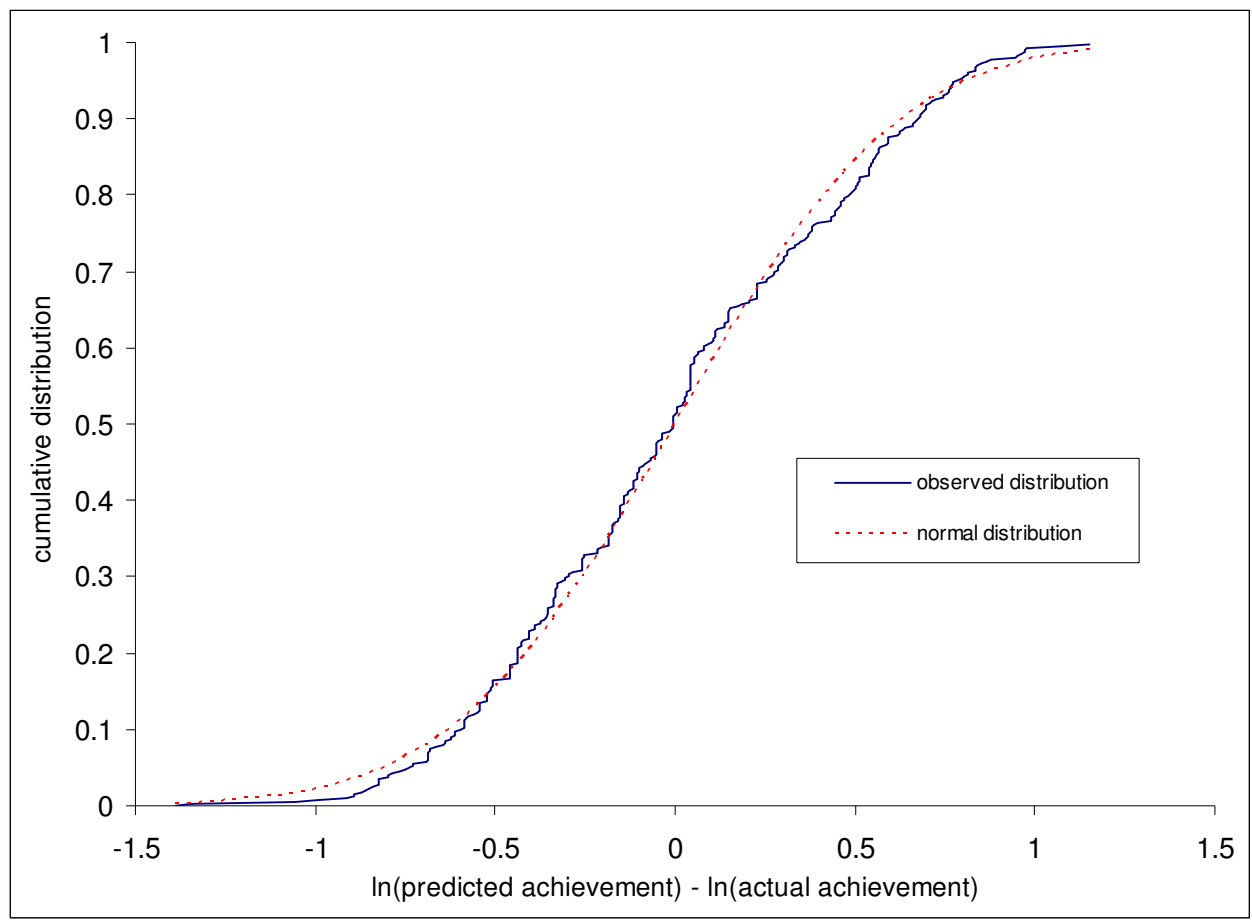

**Figure 1.** Cumulative distribution of the ratio of predicted achievement to actual achievement.

Let us now try to estimate physicist's achievement based on their fame. Table 1 shows the names of 45 pre-WWII Nobel Laureates in Physics[3], ranked according to their fame. Figure 2 shows their fame distribution. It is very similar to the fame distribution of aces (see Fig. 4 of Ref. [1]). We hypothesize that the relation between achievement and fame for physicists is, similar to aces, given by Eq. (2). A big difference with the case of aces is that we do not know the values of $\beta$ and $C$. For the case of aces, where we knew achievement values, we determined these coefficients by regression. For physicists since we do not know the achievement (we actually are to determine it) the coefficients are unknown. The fact that $\beta$ is unknown is irrelevant, as it cancels out from the ratio of achievements.

---

[3] The list includes all of the pre-WWII Nobel Laureates in Physics, excluding Charles Wilson who had so many namesakes that his fame was impossible to determine.

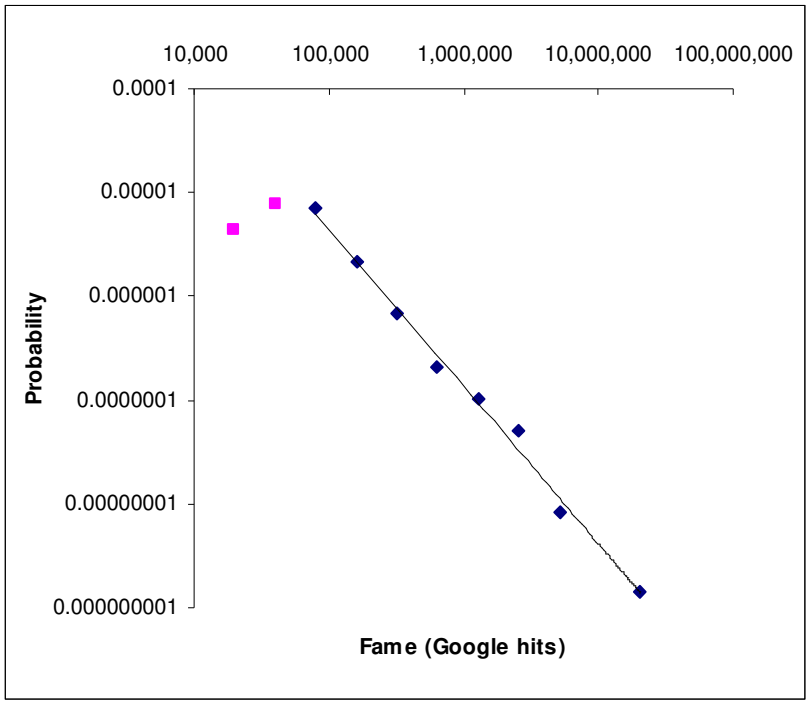

**Figure 2.** Distribution of fame of Nobel Prize winning physicists. The solid line is a power-law fit with exponent 1.5. This distribution is very similar to the fame distribution of flying aces given in Ref.[1]

The most famous physicist in Table 1 is Albert Einstein, according to Eq.(2), he is, most likely, the most achieved. Therefore, we will use him as a unit of achievement, which we denote as $A_E$. From Eq.(2) we then get:

$$\frac{A}{A_E} = \frac{\ln(F/C)}{\ln(F_E/C)} \quad (3)$$

We still need to know $C$ to find the achievement in Einsteins. While exact determination of $C$ is impossible, we can find an upper bound on it. It is the fame of the least famous person in the list: $C$ cannot be more than that because in that case the achievement of the least famous person will become negative. The least famous person on our list is Nils Dalén. His Nobel Prize is also the most contested: many believe his achievement is not worthy of it. Dalén received Nobel Prize for his invention of the automatic sun valve, which regulates a gaslight source by the action of sunlight, turning it off at dawn and on at dusk. Dalén also invented the pilot for a gas heater, which many of the contestants use in their houses. At the same time, most of the things invented by other people from our list have no practical applications, and those, which have applications, are very dangerous. Nevertheless, we will side with the contestants and assign Dalén the achievement of 0. Then we can substitute Dalén's fame, $F_D$, for $C$:

$$\frac{A}{A_E} \approx \frac{\ln(F/F_D)}{\ln(F_E/F_D)} \quad (4)$$

Eq.(4) is an estimate of a lower bound on the achievement in Einsteins. This is because $C \leq F_D$ and when $C < F_D$ Eq.(3) will give a higher value for $\frac{A}{A_E}$ than Eq.(4) for everyone but Einstein.

The estimates of achievement, computed using Eq.(4) are given in Table 1. We should note that the data presented in the table is very noisy since some physicists got additional fame for reasons other then their scientific achievement, for example for their role in public life. However, similar things happened to fighter-pilot aces that we studied in Ref.1. For example, Hermann Göring got additional web hits for his political activity. He is the second famous German WWI ace, though with his 22 victories he is only on about 60th place according to his achievement. The data shown in Fig.1 include all such cases. Let us emphasize that the error boundaries of the estimate of achievement from fame are based on the data that include all the noise and the extra hits received by aces for activities other than their career as a fighter pilot. Another objection that we encountered is that Max Planck got a lot of fame due to the singular event: renaming of Keiser Wilhelm Society into Max Planck Society.  All the institutes under auspices of the society became Max Planck institutes. Every scientific paper published by the members of Max Planck institutes automatically mentions Max Planck in its address line. Similarly, when a news article or a blog entry discusses a discovery by a member of one of the institutes, it mentions scientist's affiliation and therefore Max Planck.  Together they contribute a large share of web hits. A Google search for "Max Planck Institute" OR "Max Plank Institut" produces 6,500,000 hits. If we subtract this number from the total number of hits, we are left with 4,100,000. This shifts Max Planck from the second place to the third. The estimate of his achievement in Einsteins drops from 0.91 to 0.8 or by 12%.  The effect is thus not very big.

The estimate of achievement of every physicist listed in Table 1 (with the only exception of Dalen) is at least 15% of Einstein's achievement. For example, Dirac and Schrödinger who are 90 and 60 times less famous than Einstein appear to achieve only two times less. This may seem shocking to some people. Are these results meaningful?

Half a century ago a Nobel Prize winning physicist Lev Landau classified theoretical physicists according to their achievement using a logarithmic scale [7]. According to his ranking system, a member of the lower class achieved ten times less than a member of the preceding class. He placed Einstein in ½ class. In the 1st class he placed Bohr, Schrödinger, Heisenberg, Dirac, and Fermi[4].  Thus, he thought that Einstein contributed to Physics $\sqrt{10} \approx 3$ times more than Dirac or Schrödinger. This is close enough to our estimate, according to which Einstein achieved 2 times more than Dirac or Schrödinger. Taking into account our errors of two times more or two times less, this agreement is perfect.  Note that Landau's ranking is incomparably closer to our estimate than to a naïve estimate equating fame and achievement. The agreement becomes worse in the cases of Heisenberg and Bohr where we estimate that they achieved 0.6 and 0.7 Einsteins correspondingly. However, earlier in his life, during 1930s, Landau used another classification [7].  According to it Lorentz, Planck, Einstein, Bohr,  Heisenberg, Schrödinger, Dirac all belonged to the 1st class. Our results are compatible with this earlier Landau's classification.

A lot of recent attention was given to studies [8], where statistical analysis of very many non-expert opinions  lead to estimates agreeing with reality as good or better than expert opinions. Every webpage about a particular person expresses its creator's opinion that the person in question is worthy of it. Thus, the fact that our estimate of achievement of Nobel Prize winning physicist based on statistical analysis of numbers of webpages mentioning them agrees fairly well with expert's (Landau's) opinion may be another demonstration of wisdom of crowds[5] .

---

[4] These are the only people from Table 1, whose Landau rankings were given in [7].

[5] Surowiecki argues [8]  that one has to take an average of the guesses of a large number of people to arrive at a good estimation. In our case, many people make judgments whether to mention a particular person in their webpage. The parameter which corresponds to the average in Surowecki's book is the fraction of cases when this judgment is positive. In Ref. [1] we introduced a model were the number of people considering mentioning a particular person in their webpages is proportional to the current number of webpages mentioning the person in question. The probability that their judgment will be positive is proportional to the person's achievement. The model produced an

---

exponential growth of fame (or number of webpages) with achievement (or fraction of positive judgments). To obtain the fraction of positive judgments one has to take a logarithm of fame as the number of considerations scales exponentially with fame.

**Table 1**

| Physicist | Alternative names used in Google search, all joined using OR | June 2008 Google hits | Log over Dalen | Lower bound on the most likely achievement in Einsteins |
|---|---|---|---|---|
| ALBERT EINSTEIN | | 22,700,000 | 8.53 | 1 |
| MAX PLANCK | MAX KARL ERNST LUDWIG PLANCK | 10,600,000 | 7.77 | 0.911 |
| MARIE CURIE | | 6,300,000 | 7.25 | 0.850 |
| NIELS BOHR | | 1,890,000 | 6.04 | 0.709 |
| ENRICO FERMI | | 1,730,000 | 5.95 | 0.698 |
| GUGLIELMO MARCONI | | 1,110,000 | 5.51 | 0.646 |
| WERNER HEISENBERG | | 987,000 | 5.39 | 0.632 |
| ERWIN SCHRÖDINGER | ERWIN SCHROEDINGER | 375,000 | 4.43 | 0.519 |
| PIERRE CURIE | | 330,000 | 4.30 | 0.504 |
| WILHELM RÖNTGEN | WILHELM CONRAD RÖNTGEN WILHELM CONRAD ROENTGEN WILHELM ROENTGEN | 272,000 | 4.10 | 0.481 |
| PAUL DIRAC | PAUL ADRIEN MAURICE DIRAC PAUL AM DIRAC | 255,000 | 4.04 | 0.474 |
| LOUIS DE BROGLIE | LOUIS-VICTOR DE BROGLIE | 201,000 | 3.80 | 0.446 |
| LORD RAYLEIGH | LORD JOHN WILLIAM STRUTT RAYLEIGH | 167,000 | 3.62 | 0.424 |
| MAX VON LAUE | | 142,000 | 3.45 | 0.405 |
| HENDRIK LORENTZ | HENDRIK ANTOON LORENTZ | 119,000 | 3.28 | 0.384 |
| ROBERT MILLIKAN | ROBERT ANDREWS MILLIKAN | 112,000 | 3.22 | 0.377 |
| JAMES FRANCK | | 109,000 | 3.19 | 0.374 |
| JAMES CHADWICK | | 99,100 | 3.09 | 0.363 |
| CHARLES GUILLAUME | CHARLES EDOUARD GUILLAUME | 89,900 | 3.00 | 0.351 |
| ERNEST ORLANDO LAWRENCE | | 89,500 | 2.99 | 0.351 |
| ALBERT MICHELSON | ALBERT ABRAHAM MICHELSON | 76,600 | 2.84 | 0.333 |
| WILLIAM LAWRENCE BRAGG | | 74,500 | 2.81 | 0.329 |
| JOSEPH JOHN THOMSON | | 73,700 | 2.80 | 0.328 |
| ANTOINE BECQUEREL | ANTOINE HENRI BECQUEREL | 70,300 | 2.75 | 0.323 |
| ARTHUR COMPTON | ARTHUR HOLLY COMPTON | 66,800 | 2.70 | 0.317 |
| WILHELM WIEN | | 52,600 | 2.46 | 0.289 |
| GABRIEL LIPPMANN | | 49,300 | 2.40 | 0.281 |
| JOHANNES VAN DER WAALS | JOHANNES DIDERIK VAN DER WAALS | 48,800 | 2.39 | 0.280 |
| PIETER ZEEMAN | | 47,200 | 2.35 | 0.276 |
| WILLIAM HENRY BRAGG | | 46,800 | 2.34 | 0.275 |
| JOHANNES STARK | | 45,900 | 2.32 | 0.273 |
| MANNE SIEGBAHN | KARL MANNE GEORG SIEGBAHN | 45,000 | 2.30 | 0.270 |
| PHILIPP LENARD | PHILIPP EDUARD ANTON LENARD | 40,000 | 2.19 | 0.256 |
| CARL FERDINAND BRAUN | KARL FERDINAND BRAUN | 40,000 | 2.19 | 0.256 |
| GUSTAV HERTZ | | 37,800 | 2.13 | 0.250 |
| HEIKE KAMERLINGH-ONNES | | 35,100 | 2.06 | 0.241 |
| SIR GEORGE THOMSON | GEORGE PAGET THOMSON | 29,900 | 1.90 | 0.222 |
| CLINTON DAVISSON | CLINTON JOSEPH DAVISSON | 29,100 | 1.87 | 0.219 |
| JEAN BAPTISTE PERRIN | | 28,600 | 1.85 | 0.217 |
| CARL DAVID ANDERSON | | 26,400 | 1.77 | 0.208 |
| OWEN RICHARDSON | WILLANS RICHARDSON | 24,900 | 1.71 | 0.201 |
| CHARLES BARKLA | CHARLES GLOVER BARKLA | 24,500 | 1.70 | 0.199 |
| CHANDRASEKHARA RAMAN | CHANDRASEKHARA VENKATA RAMAN | 22,100 | 1.59 | 0.187 |
| VICTOR FRANZ HESS | | 17,200 | 1.34 | 0.157 |
| NILS DALÉN | NILS GUSTAF DALÉN NILS GUSTAF DALEN | 4,490 | 0.00 | 0 |